\documentclass[twocolumn,twocolumn]{IEEEtran}
\usepackage[T1]{fontenc}
\usepackage{color}
\usepackage{float}
\usepackage{mathrsfs}
\usepackage{amsmath}
\usepackage{amssymb}
\usepackage{graphicx}
\usepackage{cite}
\usepackage{makecell}
\usepackage{fancyhdr}
\usepackage{diagbox}
\usepackage{array}
\usepackage[raggedrightboxes]{ragged2e}
\usepackage{color}

\usepackage[unicode=true,
 bookmarks=true,bookmarksnumbered=true,bookmarksopen=true,bookmarksopenlevel=1,
 breaklinks=false,pdfborder={0 0 0},pdfborderstyle={},backref=false,colorlinks=false]
 {hyperref}
\hypersetup{pdftitle={Your Title},
 pdfauthor={Your Name},
 pdfpagelayout=OneColumn, pdfnewwindow=true, pdfstartview=XYZ, plainpages=false}

\makeatletter

\floatstyle{ruled}
\newfloat{algorithm}{tbp}{loa}
\providecommand{\algorithmname}{Algorithm}
\floatname{algorithm}{\protect\algorithmname}

\usepackage[caption=false,font=footnotesize]{subfig}
\usepackage{algorithm}
\usepackage{algorithmic}

\usepackage{multirow} 
\usepackage{amsmath} 
\usepackage{xcolor}

\allowdisplaybreaks[4]

\ifCLASSOPTIONcompsoc
\usepackage[caption=false,font=normalsize,labelfont=sf,textfont=sf]{subfig}
\else
\usepackage[caption=false,font=footnotesize]{subfig}
\fi

\usepackage{bm}
\usepackage{algorithmic}
\usepackage{algorithm}
\usepackage{graphicx}
\IEEEoverridecommandlockouts

\usepackage{lettrine}

\usepackage{geometry}
\@ifundefined{showcaptionsetup}{}{%
 \PassOptionsToPackage{caption=false}{subfig}}
\usepackage{subfig}

\makeatother

\usepackage{fancyhdr}
\begin{document}
\setlength{\parskip}{0em}
\title{\textcolor{black}{Cooperative Cognitive Dynamic System in UAV Swarms: Reconfigurable Mechanism and Framework}}

\author{\IEEEauthorblockN{Ziye Jia, \textit{Member, IEEE,} Jiahao You, Chao Dong, \IEEEmembership{Member,~IEEE,} Qihui Wu, \IEEEmembership{Fellow,~IEEE,} \\Fuhui Zhou, \IEEEmembership{Senior Member,~IEEE,} Dusit Niyato, \IEEEmembership{Fellow,~IEEE,} and Zhu Han, \IEEEmembership{Fellow,~IEEE}}
\thanks{\justifying
Ziye Jia is with the College of Electronic and Information Engineering, Nanjing University of Aeronautics and Astronautics, Nanjing 211106, China, and also with the State Key Laboratory of ISN, Xidian University, Xi’an 710071, China (e-mail: jiaziye@nuaa.edu.cn).

Jiahao You, Chao Dong, Qihui Wu and Fuhui Zhou are with the College of Electronic and Information Engineering, Nanjing University of Aeronautics and Astronautics, Nanjing 211106, China, (e-mail: yjiahao@nuaa.edu.cn, dch@nuaa.edu.cn, wuqihui@nuaa.edu, cnzhoufuhui@nuaa.edu.cn). 

Dusit Niyato is with the School of Computer Science and
Engineering, Nanyang Technological University, Singapore 639798 (e-mail:
dniyato@ntu.edu.sg).

Zhu Han is with the University of Houston, TX 77004, USA (e-mail: zhan2@uh.edu), and also with the Department of Computer Science and Engineering, Kyung Hee University, Seoul, 446-701, South Korea.}}

\maketitle
\pagestyle{headings} 
\rhead{\thepage}
\renewcommand{\headrulewidth}{0pt}
\begin{abstract}
As the demands for immediate and effective responses increase in both civilian and military domains, the unmanned aerial vehicle (UAV) swarms emerge as effective solutions, 
in which multiple cooperative UAVs can work together to achieve specific goals. 
However, how to manage such complex systems to ensure real-time adaptability lack sufficient researches.
Hence, in this paper, we propose the cooperative cognitive dynamic system (CCDS), to optimize the management for  UAV swarms. 
CCDS leverages a hierarchical and cooperative control structure that enables real-time data processing and decision. 
Accordingly, CCDS optimizes the UAV swarm management via dynamic reconfigurability and adaptive intelligent optimization. 
In addition, CCDS can be integrated with the biomimetic mechanism to efficiently allocate tasks for UAV swarms. 
Further, the distributed coordination of CCDS ensures reliable and resilient control, thus enhancing the adaptability and robustness. 
Finally, the potential challenges and future directions are analyzed, to provide insights into managing UAV swarms in dynamic heterogeneous networking.
\end{abstract}


\section{Introduction}\label{s1}
\IEEEPARstart{U}{nmanned} aerial vehicle (UAV) is widely used in multiple areas, such as surveillance, rescue, and military reconnaissance, due to the increasing matured techniques of communication, control, and manufacture.
For instance, in urban environments with overloaded networks, UAVs can be utilized as auxiliary devices to mitigate  communication burdens \cite{Bouzid_slicing_2023,Kaddour_uavcollection_2023, RDRA_Do-Duy_2021}.
Also, UAVs equipped with computing capability can also provide intelligent services via multi-access edge computing (MEC) \cite{wuwei_UAVswarm_MEC}. 
In addition, considering the general uncertainty in environments, the cooperation of UAV swarms in search and rescue operations is a significant trend \cite{ACMM_FEI_2022}.
By controlling the mobility and computational capabilities, UAVs deliver improved services, especially in remote or dynamic environments \cite{Multi_Zhao_2022}.
Besides, UAV swarms can swiftly establish communication networks, ensuring the transmission of key information during disaster or emergency cases.
However, effective coordination mechanisms within UAV swarms are necessary to ensure the  operation, and the  scheduling and deployment of UAV swarms are intractable in complex scenarios \cite{DUSF_WU_2022}.
Also, UAV swarms should share information in real-time to adapt to the variation of environments and tasks \cite{TAMW_Bai_2023}.
In brief, the challenges of  UAV swarms collaboration, including the coordination mechanisms and efficient management framework, are analyzed as follows:
\begin{itemize}    
    \item \textbf{Coordination mechanisms}: To cooperatively working in a swarm, the UAVs should firstly equipped with coordination abilities, such as the perception, inference, and  decision, and how to  coordinate these UAVs to efficiently complete tasks is intractable. Besides, how to keep autonomic learning abilities of UAV swarms to adapt to dynamic environments with risk control, is also significant for efficient and secure implementation. 
    \item \textbf{Reconfigurable framework design}: Since the UAVs in a large swarm (tens or even hundreds) work for different purposes, and are always departed into smaller swarms (subnets), it is challenging to cooperatively  manage the isomorphic UAVs in a subset for the same task, and realize coordinations among various subnets for a series of tasks. Note that  UAVs in different subnets are generally isomerous, and the possible topologies of swarms are various, a reconfigurable framework is necessary to guarantee the role transformations of UAVs within a subnet or across different subnets. 
\end{itemize} 

\begin{table*}[htpb]
	\centering 
	\caption{Characteristic comparisons of typical UAV swarms}  
	\label{table1}  

\setlength{\tabcolsep}{1.3mm}{
	\begin{tabular}{|m{3.8cm}<{\centering}|m{8cm}|m{2cm}<{\centering}|m{2cm}<{\centering}|}  
		\hline  
		{\textbf{References}}&\centering\textbf{Key contributions}&\textbf{Coordination mechanisms}&\textbf{Framework design}\\ 
        \hline
        \RaggedRight{UAV swarm based edge computing framework\cite{wuwei_UAVswarm_MEC}}&\RaggedRight{The architecture and routing protocols are proposed for the UAV swarms based edge computing.}&&\checkmark\\      
        \hline
        \RaggedRight{Cooperative search of multi UAVs\cite{ACMM_FEI_2022}}&\RaggedRight{The cooperative search strategy of multi-UAV is designed for observations in uncertain communication environments.}&\checkmark&\\     	
        \hline
        \RaggedRight{Cooperative task offloading in UAV-assisted MEC \cite{Multi_Zhao_2022}}&\RaggedRight{The cooperative multi-agent deep reinforcement learning mechanism based on UAVs are proposed for the aerial edge computing.}&\checkmark&\\  
         \hline
        \RaggedRight{Formation of  UAV swarms and avoidance strategies design\cite{DUSF_WU_2022}}&\RaggedRight{A controlling framework is proposed to integrate formation flight and swarm deployment for multiple UAVs.}&&\checkmark\\	      
        \hline
        \RaggedRight{Digital twin based cooperation of UAV swarms\cite{Lei_TICUS_2021}}&\RaggedRight{The digital twin and machine learning technologies are cooperated  to enhance the adaptability of UAV swarms.}&\checkmark&\\        
        \hline
        \RaggedRight{Path planning  for heterogeneous UAVs\cite{CPPM_Chen_2022}}&\RaggedRight{A clustering-based  path planning is proposed to improve the   coverage efficiency  for heterogeneous UAVs.}&&\checkmark\\		
		\hline
        \RaggedRight{Blockchain based monitoring system for UAV swarms\cite{Xiao_BBSCM_2021}}&\RaggedRight{The blockchain-powered  monitoring system is designed for the secure cooperation of UAV swarms.}&\checkmark&\\        
		\hline
       \RaggedRight{Scalable topology  of UAV authentication\cite{Bansal_STTOA_2022}}&\RaggedRight{Scalable topologies are designed in authentication protocols for the security and robustness of UAV swarms.}&&\checkmark\\		
		\hline
        \RaggedRight{Tracking-oriented formation control for UAV swarms \cite{UAVswarm_formationControl}}&\RaggedRight{A cooperative mechanism of tracking-oriented formation control for UAV swarms is designed based on jointly-connected topologies.} &\checkmark&\\       
		\hline  
        \RaggedRight{Cooperative mechanisms and reconfigurable framework for UAV swarms proposed in our work}&
        \raggedright{
        1)  We design the CCDS mechanism for effective implementation of UAV swarms. \\
        2)  The detailed mechanisms of intra-network and inter-network of UAV swarms are proposed.\\ 
        3)  The reconfigurable frameworks based on CCDS are designed for efficient management.}
        &\checkmark&\checkmark\\      
		\hline
	\end{tabular}}
\end{table*}    

Table \ref{table1} summarizes the key contributions  of UAV swarms in related works, from the coordination mechanisms as well as framework design.
In detail, the coordination mechanisms include the 
 cooperative searching strategy of multi-UAV for observation in uncertain communication environments \cite{ACMM_FEI_2022}, 
 cooperative task offloading mechanism for UAV-assisted MEC \cite{Multi_Zhao_2022}, 
 digital twin based cooperation to enhance the adaptability of UAV swarms  \cite{Lei_TICUS_2021}, 
 blockchain based monitoring for the secure cooperation  \cite{Xiao_BBSCM_2021}, 
 and tracking-oriented  formation control for UAV swarms  \cite{UAVswarm_formationControl}. 
 As for the framework design,  
 the MEC architecture and routing protocol \cite{wuwei_UAVswarm_MEC}, 
  controlling framework for formation flight and swarm deployment \cite{DUSF_WU_2022}, 
  clustering-based path planning for heterogeneous UAVs \cite{CPPM_Chen_2022}, 
 and a scalable topology for the security of UAV cooperation \cite{UAVswarm_formationControl}, are proposed.
 However, these works have not comprehensively considered both the coordination mechanisms and framework design, to tackle the intractable cooperation and reconfiguration in  UAV swarms with multi-subnet, heterogeneous layers, and various tasks.

Hence, in this work, we present  the cooperative cognitive dynamic system (CCDS), for collaborative decision and adaptive management of UAV swarms. 
In particular, CCDS can facilitate efficient management of UAV swarms via cooperative perception, decision,  execution, and reconfiguration. 
We first propose an overview of CCDS, as well as the detailed mechanisms in both intra-network and inter-network of UAV swarms. 
Then, we analyze the abilities of UAV swarms equipped with CCDS and artificial intelligence (AI) techniques. 
Further, the mechanisms of dynamic reconfiguration, biomimetics, and reliable control of UAV swarms with CCDS are proposed. 
Moreover, we discuss the applications and provide a case study to verify the performance of CCDS in UAV swarms. 
Finally, the challenges and directions are analyzed. The key contributions of this work are summarized as:

\begin{itemize}
\item We present the CCDS mechanisms for efficient management of UAV swarms, based on the techniques of cooperative attention perception,  learning and inference, and risk control.
\item The  mechanisms in both intra-network and inter-network of UAV swarms based on CCDS are designed, to enable the  operation  efficiency. 
\item The reconfigurable frameworks for UAV swarms related with CCDS are proposed, including the biomimetics for heterogeneous UAVs, and the reliable control mechanism.
\end{itemize}

\begin{figure*}[!t]
    \centering
    \includegraphics[width=18.2cm]{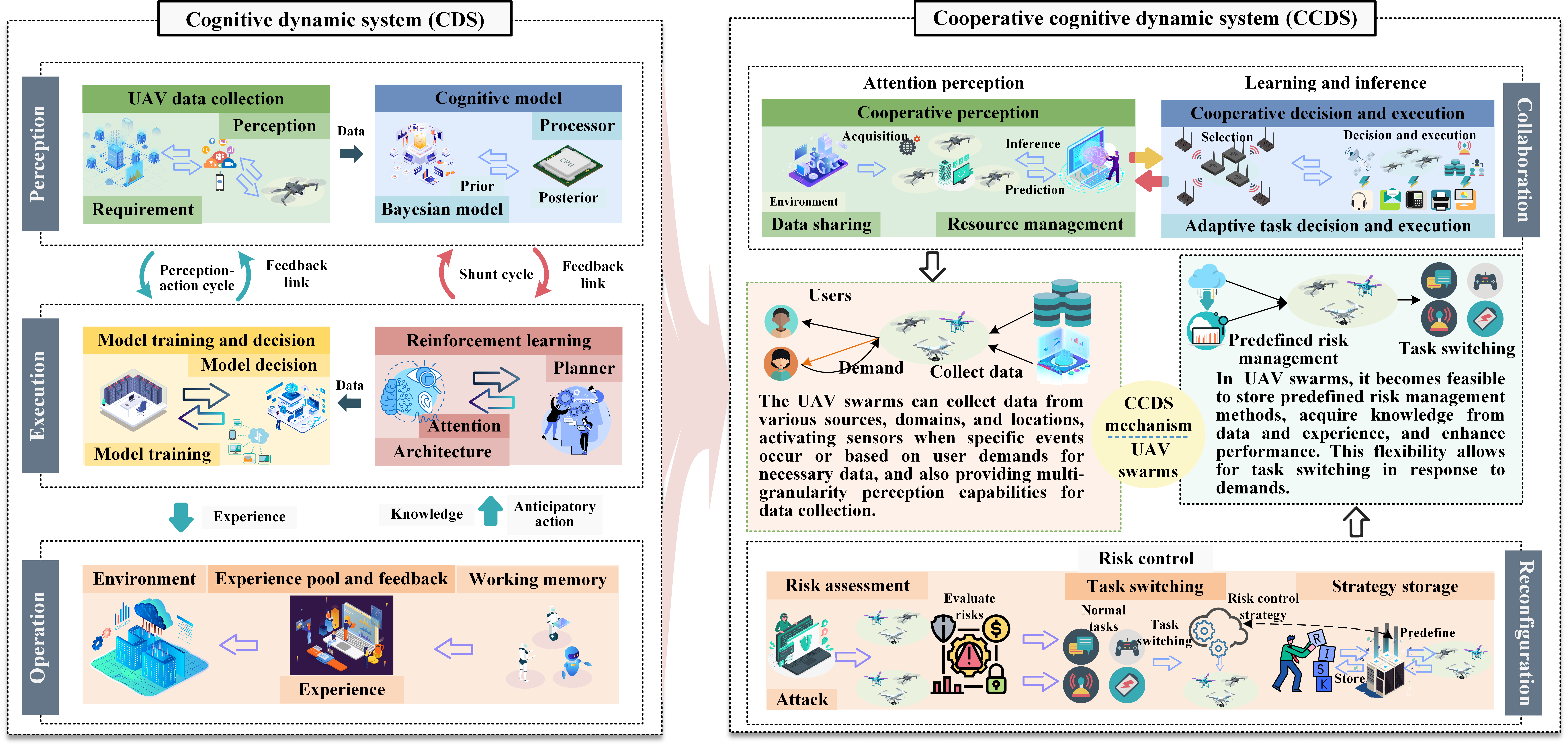}
    \caption{The evolution and mechanism of CCDS, including modules of attention perception, learning and inference, and risk control.} 
    \label{f2}
\end{figure*}
 
\section{Overview of Cooperative Cognitive Dynamic System}\label{s2} 
To enhance the collaborative decision and adaptive management of UAV swarms, we present the CCDS mechanisms, as shown in Fig. \ref{f2}. 
Generally, the traditional cognitive dynamic system (CDS) is a cognition model, drawing inspiration from concepts  in neuroscience and human brain cognition \cite{Cognitive_Hilal_2023}.
CDS includes modules of  perception, execution, and  operation, in which the perception-action cycle and  shunt cycle \cite{Haykin_CDSBC_2016} are significant. 
In detail, UAVs perceive the environment via sensors and execute based on the extracted information,
and then, the results are sent back to the perception module, completing a perception-action cycle.
Besides, the  shunt cycle operates in the reverse direction, which involves re-perceiving based on execution information, and  provides feedback to the execution module for further action.
After the execution phase, experiences are transformed into operations for extracting knowledge. Anticipatory actions within operations assist in refining the execution and creating a loop for continuous improvement.
However, the CDS design has not considered the cooperation, which cannot be directly used by UAV swarms.

Therefore, we propose CCDS to enable the intelligent and collaborative management for UAV swarms.
CCDS integrates AI technologies, such as multi-agent reinforcement learning, to  enhance decision and promote efficient communication among UAVs. 
In addition, CCDS focuses on real-time intelligent data sharing and interaction,  enabling dynamic resource management, adaptive task execution, and resilience of UAV swarms. 
Besides, it aims to improve the operational efficiency and adaptability of UAVs in dynamic environments, such as disaster relief, surveillance, military reconnaissance, and public safety missions.
In detail, CCDS is composed of three parts: the attention perception, learning and inference, and risk control. 

\subsubsection{Attention perception}
The attention perception follows specific principles, including the semantic information measurement, which enables the system to measure the relevance and importance of the information. 
Besides, the perception depends on demand, and the system collects data when necessary.
Furthermore, the process involves analyzing information  from multi-domain, such as geography, environment, and demands, which enables the system to allocate perception effectively and concurrently monitor different areas. 

\subsubsection{Learning and inference}
The learning and inference module  indicates that the system continuously updates knowledge and adapts based on new information. 
It also includes prediction, i.e., the system can anticipate future events or states. 
Additionally, the process encompasses deduction, allowing the system to make logical conclusions from available information. 
Besides, it involves internal self-inference,  refering to the system ability to draw conclusions based on internal knowledge.
\subsubsection{Risk control}
The risk control module is crucial for efficient operation of UAV swarms and task implementation in dynamic environments with uncertainty.
The detailed strategies encompass risk assessment, task switching, and strategy storage. 
In detail, risk assessment involves the system capability to continuously evaluate potential risks. 
Task switching refers to the  ability to change current tasks in response. 
The strategy storage indicates the system capability to store and retrieve predefined risk control approaches.

\begin{figure*}[!t]
    \centering
    \includegraphics[width=18.3cm]{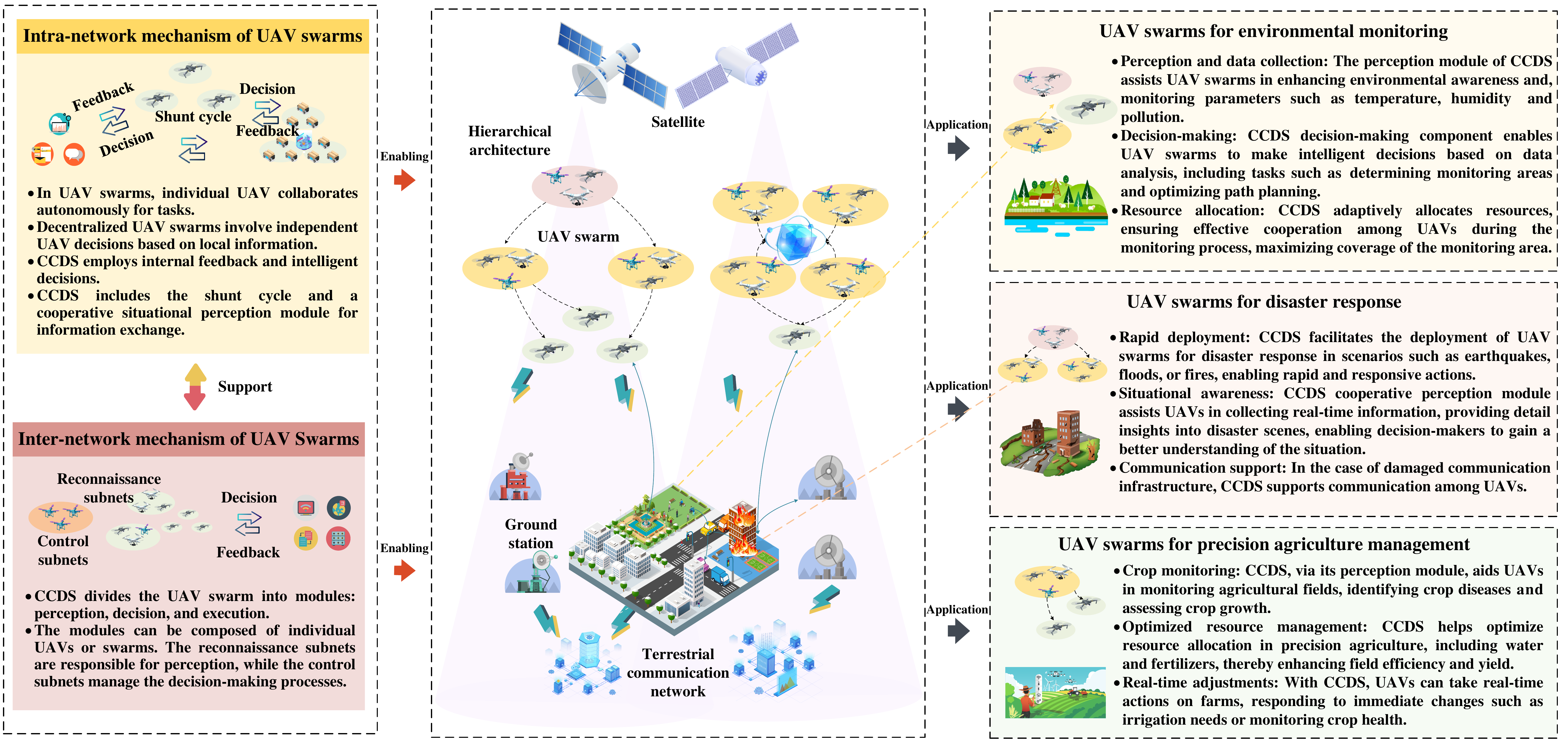}
     \caption{UAV swarms with CCDS: the intra-network and inter-network mechanisms, as well as the operational procedures in typical applications.}
    \label{f1}
\end{figure*}

In addition, there are two general types of decision mechanisms for CCDS. 
The first is {\em independent decision}, where each module makes decisions sequentially and separately.
Although the approach is simple to implement, it requires a predefined order and synchronization, which limits the collaboration among UAV swarms.
Another approach is {\em collaborative decision}, which combines the solution spaces of each module into new decision variables. 
Then, tasks such as decisions of UAV spectrum access and routing are jointly considered, leading to potential improvements in the  decision-making quality.
However, the approach causes the problem of combinatorial explosion and communication overhead, increasing the decision complexity.
To address this issue, by integrating machine learning and optimization techniques for adaptive strategy searching, CCDS can refine the solution space to alleviate the challenge of combinatorial explosion.

As a summary, the CCDS mechanism is based on attention perception, learning and inference, and risk control, to realize collaborative decisions,  which can help UAV swarms to implement with intelligence, robustness, and efficiency.

\section{CCDS Facilitated UAV Swarms}\label{s3}   

\subsection{Intra-network Mechanism of UAV Swarms}
In UAV swarms, as an autonomous agent, an individual UAV can engage in the multi-agent cooperation to execute complex tasks. Traditionally, UAV swarms operate under a centralized control structure, and the central node generates the actions of all UAVs. 
However, the mechanism is limited by the center node failure and communication bottlenecks.
In contrast, in the decentralized UAV swarm, each UAV independently makes decisions based on local information and interactions with neighboring UAVs. 
Hence, without relying on a central controller allows for flexibility and adaptability, since individual UAVs can adjust actions in real-time to suit the environment. 
In detail, the decentralized communication approach empowers UAVs to share data and synchronize decisions in real-time via direct UAV-to-UAV communications.
The integration allows the system to learn from past experiences, and adjust decision accordingly. 
Therefore, CCDS enables efficient data sharing, collaborative reasoning, and adaptive resource allocation,  enhancing the decision efficiency and adaptability within UAV swarms.

CCDS also incorporates the shunt cycle, which serves as the internal mechanism in UAV swarms and manages information exchange among  networking functions.
Then, the cooperative situational perception module utilizes discrete state acquisition, global situational inference, and evolution trend prediction for UAV swarms. 
In detail, the perception process provides information for the decision and execution module via the shunt cycle. 
Besides, each functional module is equipped with the shunt cycle, which directly communicates with the perception module for instant feedback or updated results, ensuring the information accuracy. 
Specifically, the shunt cycle facilitates a learning mechanism similar to the experience replay technique in reinforcement learning, allowing the system to learn from past experiences, such as navigating through complex environments.
By integrating historical data into current decision-making processes, the UAV swarm can optimize its strategies based on previous outcomes.

For example, in the UAV swarm monitoring a forest for wildfires, as shown in Fig. \ref{f1}, the modules work as:
\vspace{-0.3cm}
\subsubsection{Information collection}
UAVs collect data such as temperature, humidity, and smoke density, and share with the swarm.
\subsubsection{Situational perception}
The system assesses the forest status such as normal, potential fire, or confirmed fire.
\subsubsection{Decision and feedback}
The decision module obtains the information and identify the potential  threat, and then decides if the risk is continuously monitored.
\subsubsection{Instant communication}
Each module uses shunt cycles for real-time feedback. For example, if a UAV senses a spike in temperature or smoke, it alerts the swarm for prompt action.
\subsubsection{Adaptability}
If a risk is detected, the swarm may shift its mission to focus on fire monitoring. Environmental variations trigger updates.

\subsection{Inter-network Mechanism of UAV Swarms}
A heterogeneous UAV swarm generally operates on a hierarchical structure and the swarm is organized into different levels of decision, i.e., various subnets have distinct roles and responsibilities. 
For example, reconnaissance subnets focus on data collection (perception), control subnets handle decision-making processes, and strike subnets are responsible for task execution. 
However, with the growing number of  tasks, a single homogeneous UAV swarm may fail to meet the demands. 

CCDS introduces an adaptive communication strategy (ACS), enabling a dynamic and responsive interaction among the diverse functional modules of UAV swarms, allowing for continuous feedback and subsequent decision modification. In detail, ACS is based on the principle of dynamic network configuration, where communication links adjust in real-time according to the task priority, data volume, and network congestion levels.
Additionally, CCDS partitions the UAV swarm into different functional modules: perception, decision, and execution. 
The modules can be flexibly composed of different subnets. 
Also, multiple subnets in UAV swarms can be merged into a superior unit for intensive demands.

Since tasks have different priorities based on specific requirements, assigning high priority to control signals is essential to maintain continuous and uninterrupted communication.
Moreover, due to the presence of heterogeneous subnets with varying requirements in terms of routing, perception, and decision, it is necessary to adjust the weights of tasks for different UAVs during the cooperation process. 
For instance, UAVs equipped with enhanced perception abilities should be allocated with high weights in perception tasks, to ensure more reliable and accurate data analysis within the UAV swarm. 
As for the  UAVs with special tasks, the function adjustment among different subnets should be careful.

\subsection{Abilities of UAV swarms with CCDS}
Fig. \ref{f1} provides a scenario of UAV swarms with CCDS, including the intra-network and inter-network mechanisms, as well as  the operational procedures in typical applications. 
In detail,  UAV swarms are equipped with devices of communication, computing, storage, etc. 
Consequently, UAVs are equipped with the capabilities of perception, storage, transmission, and computation to accomplish complex tasks, such as autonomous environmental monitoring, disaster response, and precision agriculture management.
With CCDS, one subnet is designed for a specific task, and each has a different network topology. In particular, UAV swarms energized with CCDS and AI have the following abilities:

\subsubsection{Advanced perception}
Sensor fusion techniques integrate data from various sensors, including cameras, LiDAR, and infrared cameras, providing a comprehensive understanding of the environment.
By intelligently processing sensor data and employing advanced signal processing and machine learning algorithms, CCDS enhances the perception capabilities of UAV swarms. 

\subsubsection{Autonomic decision}

The application of AI in the UAVs brings  effective autonomic decisions.
For instance, reinforcement learning enables UAV swarms to independently plan flight trajectories and navigate in complex environments.
Further, UAVs leverage technologies such as computer vision  and advanced communication systems,  to recognize obstacles and ensure safe  mission execution.

\subsubsection{Augmented adaptability}
Due to the AI-driven predictive models and real-time data analysis, UAV swarms can dynamically adapt to the variation of environments and demands. 
The ability for dynamic adaptation is driven by advanced machine learning algorithms, such as  deep reinforcement learning.

In short, with the assistance of CCDS, UAV swarms can flexibly implement in both intra-network and inter-network, with advanced perception, autonomic decision, and augmented adaptability. 

\begin{figure*}[!t]
    \centering
    \includegraphics[width=18cm]{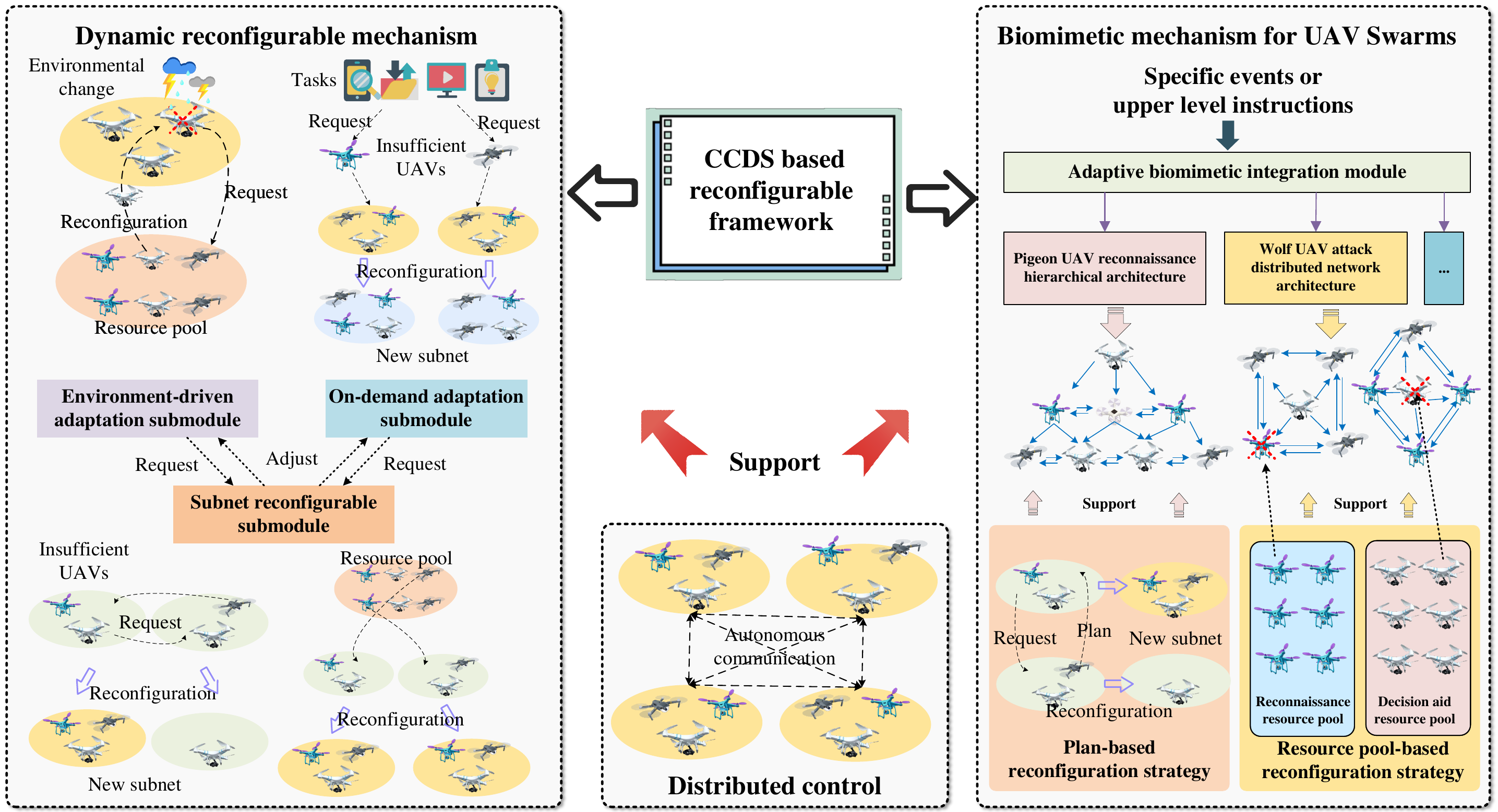}
    \caption{CCDS based reconfigurable framework, which is supported by the distributed control mode, and implement according to the dynamic reconfigurable mechanisms and biomimetic mechanisms of UAV swarms.}
    \label{f3}
\end{figure*}

\section{CCDS based Reconfigurable Framework}
\subsection{Dynamic Reconfigurable Mechanism}
To realize the dynamic reconfiguration of UAV swarms, we design the CCDS based reconfiguration framework, as shown in Fig. \ref{f3}.
The framework includes the dynamic reconfigurable mechanism and the biomimetic mechanism, and supported by the distributed control mode.
In detail, the dynamic reconfigurable mechanism allows UAV swarms to  change the configuration in real-time, enabling flexible response to task variation.
When  CCDS identifies a demand, or the upper-layer application proactively triggers the network, the reconfigurable submodule initiates the network reconfiguration. 
In particular, the submodules are detailed as follows.

\subsubsection{Environment-driven adaptation submodule}
Due to dynamic environmental changes, it is crucial for reconfiguration to ensure stability of UAV swarms. 
For instance, in scenarios where environmental interference affects communication links, the system can autonomously adapt by reconfiguring the network to maintain a robust connection.

\subsubsection{On-demand adaptation submodule}
The primary function involves the integration and disintegration of UAVs into swarms based on demands. 
For instance, if a task is assigned to a UAV, but the capability of the single UAV is insufficient to handle the task, a request can be made to other UAVs to form  a  UAV swarm, to complete the  requirements.

\subsubsection{Subnet reconfigurable submodule}
The subnet reconfigurable submodule refers to   the UAV reconfiguration from the resource pool or other swarms, to   constitute  new subnets for different tasks. 
For example, in Fig. \ref{f3}, when a UAV swarm lacks resources, it can initiate reconfiguration by drawing other UAVs from the resource pool.

The on-demand adaptation submodule is central for task allocation to ensure  service stability,
and it interacts with the subnet reconfigurable submodule to reorganize resources or create new subnets, ensuring efficient task execution. 
Simultaneously, the environment-driven adaptation submodule monitors environmental changes autonomously, and  prompts the subnet reconfigurable submodule to adjust topologies.
Therefore, by providing the flexibility and resilience, the reconfigurable modules enhance the efficiency and reliability of UAV swarms,  particularly in complex and challenging environments.

\subsection{Biomimetic Mechanism for Heterogeneous UAV Swarms}
Based on different tasks, UAV swarms work in specific topologies, as shown in Fig. \ref{f3}. 
Biomimetic mechanisms enable UAV swarms to dynamically adapt to different environments and mission parameters, similar to the adaptability in biological swarms.	
Hence, the adaptability is crucial for maintaining operations in the face of hardware malfunctions or uncertain requirements.	
In detail, by imitating the dynamic flocking behavior of pigeons, UAV swarms can efficiently navigate and cover extensive areas with limited resources, thereby enhancing the surveillance efficiency.	
In addition, adopting the coordinated attack strategies of wolves provides powerful strike capabilities.	
These strategies can enhance the operational efficiency and flexibility of UAV swarms, and improve the real-time reconfigurability, ensuring superior performance across diverse tasks and conditions.
Therefore, the distinct biomimetic swarm formations correspond to unique network topology requirements. 

Meanwhile, it is crucial to dynamically reconfigure the network topology, to facilitate the fusion of different biomimetic swarm formations, as illustrated in Fig. \ref{f3}.
Besides, we propose two distinctive reconfiguration strategies with the CCDS framework, i.e., the {\em plan-based strategy} and the {\em resource pool-based strategy}.

\setcounter{subsubsection}{0}
\subsubsection{Plan-based reconfiguration strategy}
The plan-based reconfiguration strategy is suitable for scenarios where subnets can complement each other by partially merging.
After reconfiguration, the previous subnets are combined to form a new operational subnet. 
However, the strategy  is limited by waiting for compatible subnets to emerge, which may result in a long delay.
Additionally, during the reconfiguration process, networks should switch to self-organization mode before altering the network topology.

\subsubsection{Resource pool-based reconfiguration strategy}
The resource pool-based reconfiguration strategy is characterized to maintain a swarm resource pool. 
When a subnet becomes unable to fulfill missions, the nodes are reassigned to the resource pool. 
Once the resource pool accumulates adequate nodes to satisfy the mission requirements, a new subnet can be reconfigured. After forming a subnet, UAVs designate the gateway nodes and establish a connection with the resource pool. 
When the mission is completed, the 	UAVs are released to the resource pool. 

After completing a task such as rescue or strike, heterogeneous subnet groups can apply for rejoining the UAV resource pool. 
It highlights the dynamics  during missions, which necessitates an adaptive process to modify the network topology. 
Therefore, the network topology  varies based on different biomimetic swarm formations, and adjusts when rejoining the resource pool. The process is detailed as:
\begin{enumerate}
    \item A command from the control center instructs the UAV swarm to return to the resource pool.
    \item The UAV swarm receives the returned request, and designates a couple of nodes as gateway.
    \item The gateway acquires topology information and architecture of the connected swarm,  and decides which pool for each UAV to return.
    \item The gateway issues commands to the UAV swarms, leading to the topology variation and UAVs entering into the resource pool.
\end{enumerate}

\begin{figure*}[!t]
    \centering
    \includegraphics[width=18cm]{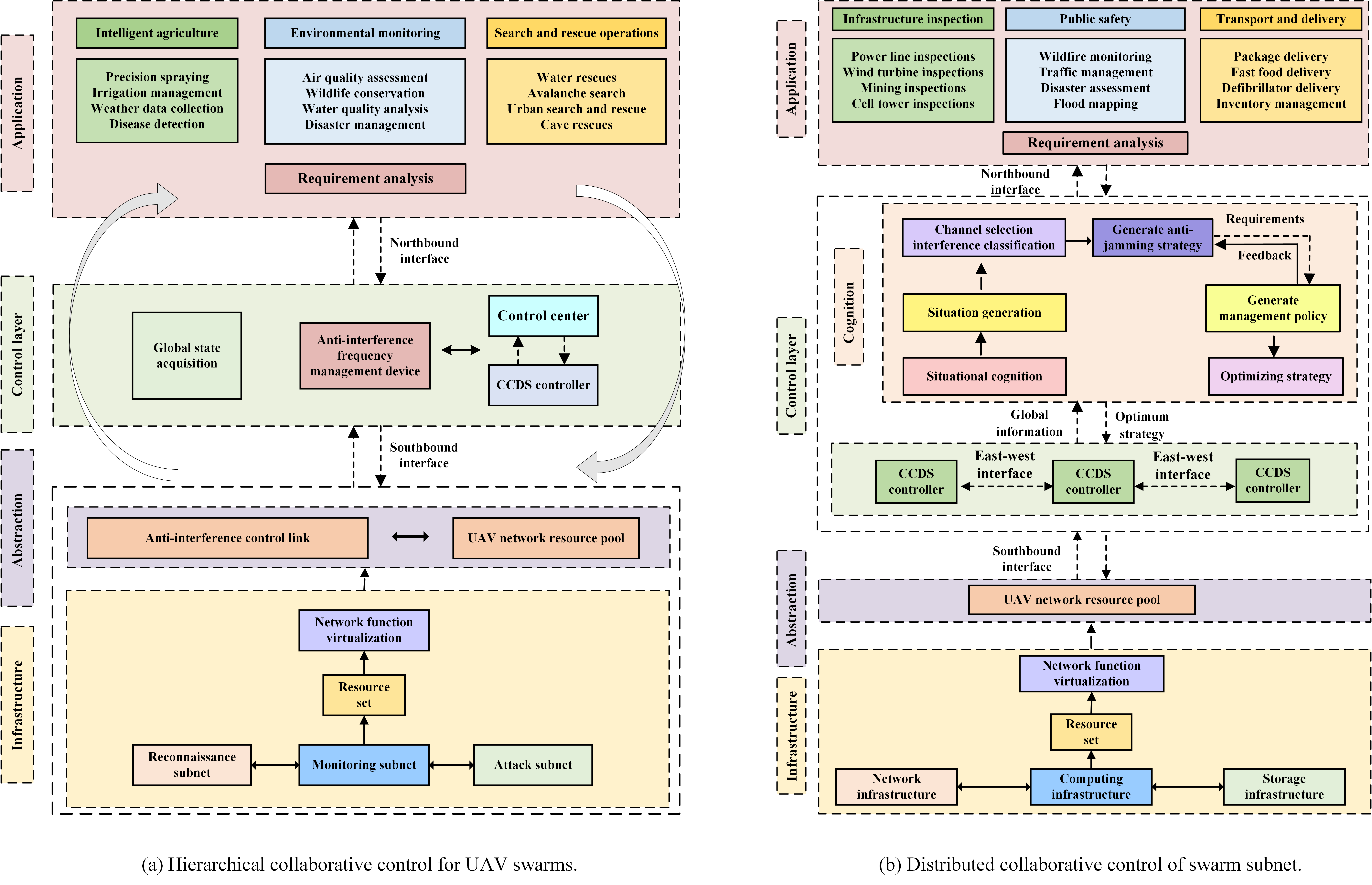}
    \caption{Risk control mechanisms for UAV swarms: hierarchical collaborative control for UAV swarms and distributed collaborative control for subnets.}
    \label{f5}
\end{figure*}

\subsection{Risk Control Mechanism}
Since UAV swarms operate in dynamic and unpredictable environments, the reliable coordination and risk control are crucial to ensure successful reconfiguration. 
The  framework comprises multi layers of infrastructure, abstraction, CCDS control, and applications, as shown in Fig. \ref{f5}.

\subsubsection{Infrastructure layer}
The infrastructure layer provides essential resources of UAV swarms, including reconnaissance subnets, monitoring subnets, attack subnets, etc. Moreover, the network function virtualization technique is used for resource management.

\subsubsection{Abstraction layer}
This layer can conceal infrastructure complexities and provide high-level interfaces. 
Besides, the anti-interference control link is applied to guarantee the transmission of control information in UAV swarms, which can protect the control signals from being disrupted by external electromagnetic interference.
\subsubsection{Control layer}
The layer works by integrating perception, decision, and execution mechanisms. 
It also encompasses the anti-interference frequency decision and an additional cognition layer. 
In detail, the cognition layer provides an additional level of sophistication, facilitating the development of optimizing strategies.

\begin{figure}
    \centering
    \subfloat[The average execution time.]{\includegraphics[width=0.8\linewidth]{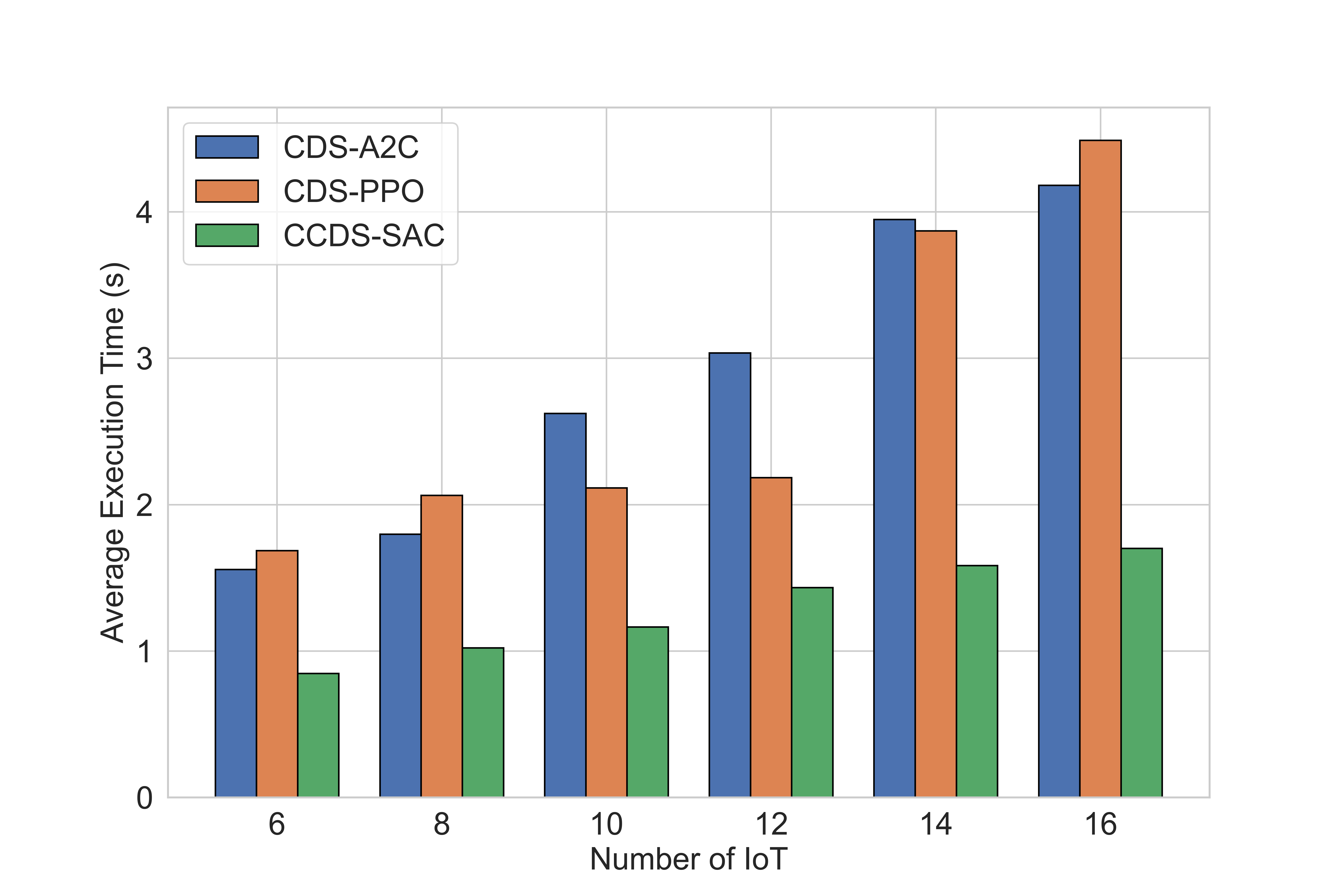}\label{f5a}}\\ 
    \subfloat[The average computation rate.]{\includegraphics[width=0.8\linewidth]{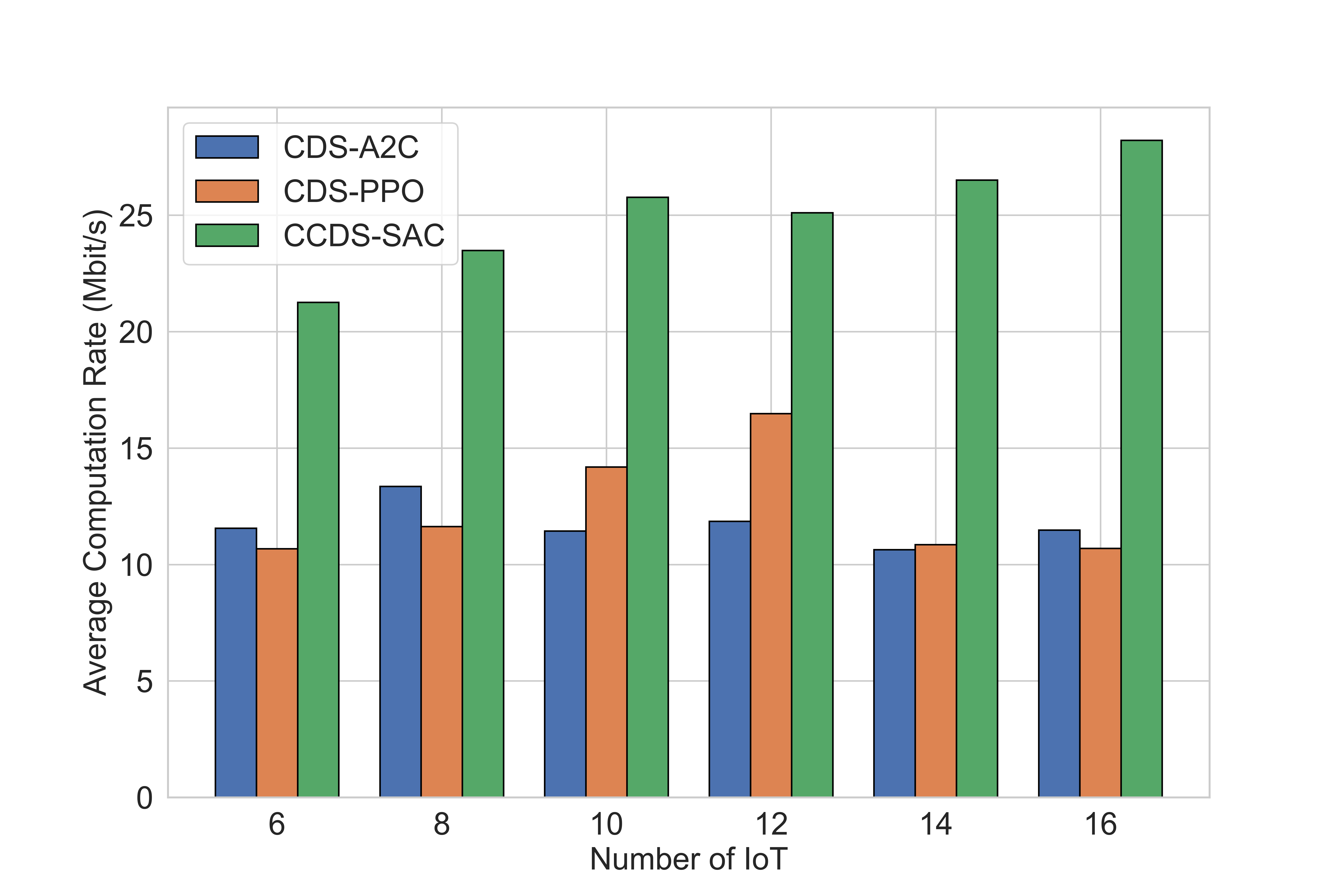}\label{f5b}}\\
    \subfloat[The average offloaded data.]{\includegraphics[width=0.8\linewidth]{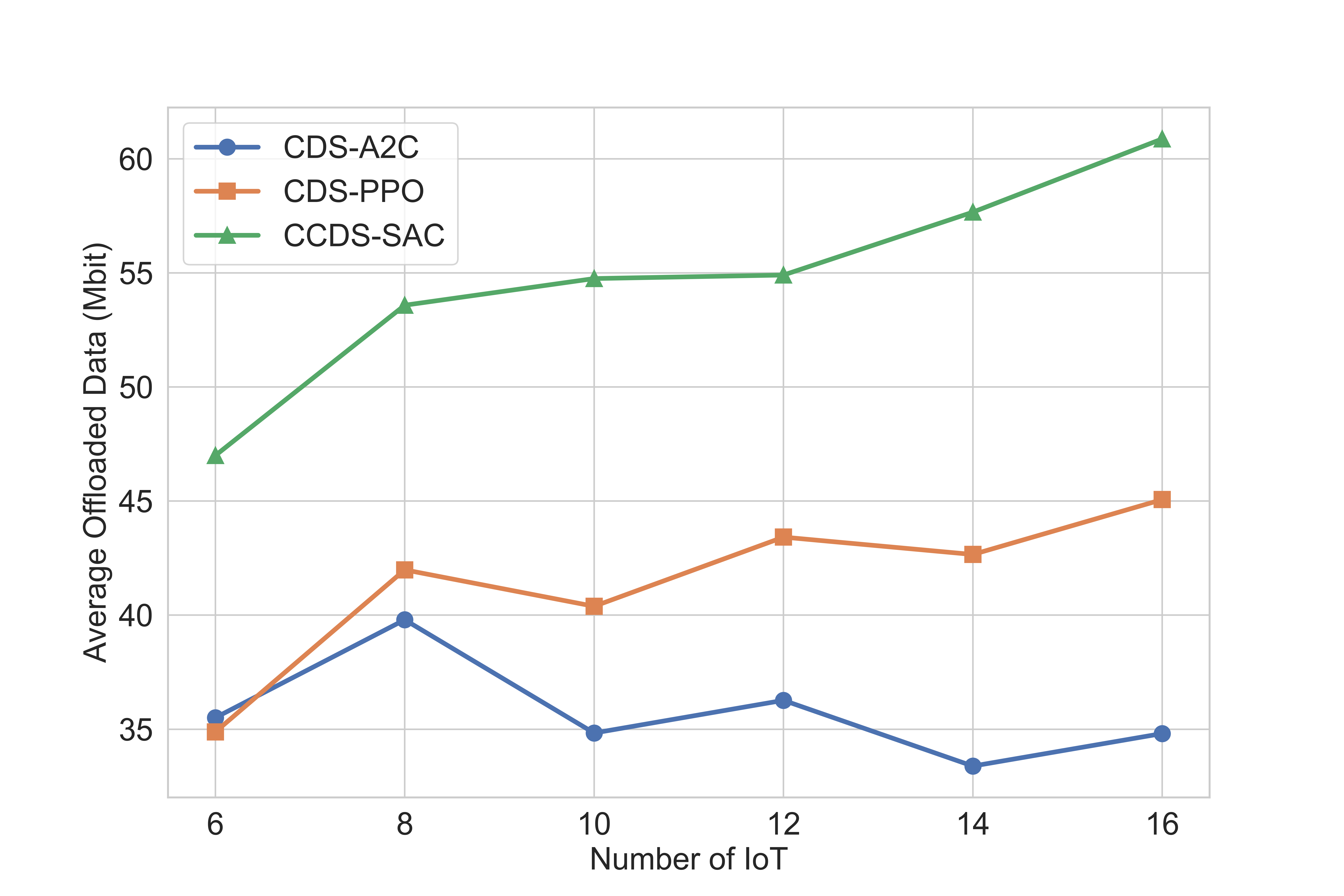}\label{f5c}}\\ 
    \caption{Numerical results for task offloading.}
    \label{fig5}
\end{figure}
\subsubsection{Application layer}
This layer involves various UAV applications, such as intelligent agriculture, environmental monitoring, infrastructure inspection, etc. 
The application layer works by translating task requirements into commands to meet the demands of users and decision-makers. 

Specifically, Fig. \ref{f5}a illustrates the hierarchical collaborative control for UAV swarms.
By leveraging distributed computation, UAVs can share information, make decentralized decisions, and adjust responses based on feedback. 
The mechanism integrates centralized control with decentralized execution, fostering a dynamic operational framework. The centralized control ensures that UAV swarms are guided by objectives, thus enhancing the coherent operational direction. 
Concurrently, the decentralized execution allows individual UAVs to autonomously respond to immediate environmental challenges. The mechanism optimally balances UAV swarms and the autonomy of individual UAVs, ensuring an effective balance between the global objectives and local responsiveness.

The control layer is the centre to ensure effective data transmission and cooperation, which includes global state acquisition, anti-interference frequency management device, control center and CCDS controller.
Besides, via the south-north interfaces, the control layer manages the information flow, to ensure the data transmission and cooperation. 
The control center effectively coordinates the activities of all UAV swarms, enabling the global information for each UAV. 
The CCDS controller is mainly responsible for the management of different subnets.
Besides, the interaction between the control center and CCDS controller enables UAVs to obtain comprehensive understanding of the environment. 
For instance, UAVs for wildfire monitoring tasks dynamically share and respond to the fire spread data, and adjust their paths to decrease risks.

In Fig. \ref{f5}b, the distributed collaborative control is used for swarm subnets.
In detail, information exchange and decision occur via the east-west interfaces among UAVs in the control layer.
By leveraging CCDS, UAVs contribute to collective intelligence by  perceiving the environment and the actions of neighboring UAVs, and autonomously generating management strategies.
In scenarios with rapid variation of risks and unpredictable environment, such as the disaster  or battlefield, it empowers the UAV swarm with rapid adjustment via the dynamic decision-making mechanisms and adaptive strategies. 
Specifically, by utilizing distributed control mechanisms, UAVs in a disaster zone can share real-time data to navigate and reduce risks, enhancing response strategies for safety and efficiency.
In conclusion, a reconfigurable framework based on CCDS is developed for UAV swarms, including dynamic reconfigurable mechanisms and biomimetic mechanisms, with risk control to ensure the stability of UAV swarms.

\section{Applications and case study}
Facing the natural disasters such as earthquakes or hurricanes, traditional  response and rescue operations are limited by inaccessible terrain, resources, and time. 
To overcome these challenges, UAV swarms with CCDS can be applied due to the following properties.

\subsubsection{Swarm formation and deployment}
UAV swarms are deployed in disaster areas, and UAVs are equipped with cameras, communication modules, etc. 
Then, UAVs autonomously assemble into a swarm, establishing vital communication links for disaster recovery.
\subsubsection{Intelligent perception and mapping}
UAV leverages sensors and cameras to obtain real-time data in the disaster area. 
Then, AI algorithms can be applied to analyzing data, identifying survivors, assessing damages, and creating maps.
\subsubsection{Search and rescue operations}
AI-powered algorithms can sort the priority of  rescue areas based on detected signs of life and structural instability.
Then, UAV swarms can autonomously navigate in the disaster area avoiding obstacles. 
\subsubsection{Real-time monitoring}
UAV swarms can continuously monitor the disaster area, detecting any changes in conditions or emerging risks. 
Then, UAVs send real-time information to the ground center, allowing decision-makers to promptly adopt response strategies.
\subsubsection{Data analysis and decision support}
With UAV swarms, the collected  environmental data, such as images, are efficiently transmitted to the command center. 
Then, AI algorithms can process the data to extract valuable insights. 

\textbf{Case study:}
Internet of things (IoT) devices collect real-time environmental and operational data to UAV swarms. Then, UAVs serving as mobile processors, handle the data and respond to environmental changes by making decisions. 
In detail, we conduct numerical simulations in a scenario involving 6-16 IoT devices, 6 UAVs distributed within a 1km$\times$1km area, where IoT devices generate tasks that are offloaded to UAVs for execution, to verify the performance of CCDS applied to task offloading. In detail, CDS employs a non-cooperative approach, i.e., each UAV works independently. 
In contrast, CCDS promotes cooperations among UAVs via collaborative communication and mutual offloading, to optimize task offloading and execution.
Additionally, we assume that CCDS requires the utilization of advanced algorithms. Consequently, the soft actor-critic (SAC) algorithm is implemented in CCDS, while the proximal policy optimization (PPO) and advantage actor-critic (A2C) algorithms serve as benchmarks in CDS.

In Fig. \ref{f5a}, the reduced average execution time under CCDS indicates that tasks are completed more swiftly, attributable to the combined computational power and shared intelligence.
Fig. \ref{f5b} shows a higher computation rate, indicating that more tasks are successfully completed within a given time frame under the CCDS framework. 
The increment of offloaded data shown in Fig. \ref{f5c} represents more tasks being offloaded, which can prevent UAV swarms from becoming a bottleneck due to overloading. 	
Therefore, CCDS is an effective solution for optimizing task execution in UAV swarms.

\section{Challenges and Directions}
CCDS brings a series of challenges and research issues such as system complexity, adaptability, and interoperability.
By tackling these challenges, UAV swarms in dynamic environments will be improved, enhancing the efficiency of CCDS for various future applications.

\subsection{Challenges and Research Issues}
\subsubsection{System complexity}
The CCDS integration of varied subnets presents implementation challenges, such as interoperability demands, infrastructure reliance, managing complexity and energy optimization. Hence, developing a straightforward, easy-to-implement model with efficiency is a significant issue.

\subsubsection{Adaptability} 
In the varying environment, the navigation of UAV swarms faces difficult factors, such as real-time fluctuations in environments, and burst tasks.
Hence, in these scenarios, UAV swarms need to quickly adapt to the new information and evolving situations.

\subsubsection{Energy consumption}
UAV energy consumption is influenced by payloads, flight dynamics, and environmental conditions. 
The wind, air density, temperature, and humidity considerably affect propulsion energy demands and aerodynamic efficiency, highlighting the need for adaptive energy management strategies.

\subsection{Future Directions}
\subsubsection{Advancements with AI}
Utilizing advancements with AI can significantly reduce the complexities of integrating heterogeneous subnets within the CCDS, thereby improving interoperability and optimizing energy management.

\subsubsection{Network architecture improvements} 
It is necessary for more researches to improve network architectures, specifically focusing on ensuring seamless interoperability among heterogeneous subnets.

\subsubsection{Security and privacy} 
With improved autonomy and cognition in UAV swarms, the security and privacy of data becomes vital. Future  explorations should develop robust security protocols for the safe operation in UAV swarms.

\section{CONCLUSION}
In this work, we have proposed the CCDS to address  reconfigurable mechanisms and frameworks for  the UAV swarm management.
The evolution and mechanism of CCDS has been presented to illustrate the insights and advantages in UAV swarms, and the abilities of UAV swarms with CCDS have been analyzed. 
Then, we have detailed the CCDS based reconfigurable framework, to facilitate the flexible coordination of UAV swarms.
Besides, the possible applications and case study have be presented, and the challenges and directions have been analyzed finally.  We hope the future researches can  gradually improve the open issues.


\bibliographystyle{IEEEtran}
\bibliography{references}

\begin{thebibliography}{10}
\providecommand{\url}[1]{#1}
\csname url@samestyle\endcsname
\providecommand{\newblock}{\relax}
\providecommand{\bibinfo}[2]{#2}
\providecommand{\BIBentrySTDinterwordspacing}{\spaceskip=0pt\relax}
\providecommand{\BIBentryALTinterwordstretchfactor}{4}
\providecommand{\BIBentryALTinterwordspacing}{\spaceskip=\fontdimen2\font plus
\BIBentryALTinterwordstretchfactor\fontdimen3\font minus
  \fontdimen4\font\relax}
\providecommand{\BIBforeignlanguage}[2]{{%
\expandafter\ifx\csname l@#1\endcsname\relax
\typeout{** WARNING: IEEEtran.bst: No hyphenation pattern has been}%
\typeout{** loaded for the language `#1'. Using the pattern for}%
\typeout{** the default language instead.}%
\else
\language=\csname l@#1\endcsname
\fi
#2}}
\providecommand{\BIBdecl}{\relax}
\BIBdecl

\bibitem{Bouzid_slicing_2023}
T.~Bouzid, N.~Chaib, M.~L. Bensaad, and O.~S. Oubbati, ``{5G} network slicing
  with unmanned aerial vehicles: Taxonomy, survey, and future directions,''
  \emph{T EMERG TELECOMMUN T}, vol.~34, no.~3, p. 4721, Dec. 2023.

\bibitem{Kaddour_uavcollection_2023}
K.~Messaoudi, O.~S. Oubbati, A.~Rachedi, A.~Lakas, T.~Bendouma, and N.~Chaib,
  ``A survey of {UAV}-based data collection: Challenges, solutions and future
  perspectives,'' \emph{J NETW COMPUT APPL}, vol. 216, p. 103670, May 2023.

\bibitem{RDRA_Do-Duy_2021}
T.~Do-Duy, L.~D. Nguyen, T.~Q. Duong, S.~R. Khosravirad, and H.~Claussen,
  ``Joint optimisation of real-time deployment and resource allocation for
  {UAV}-aided disaster emergency communications,'' \emph{IEEE J. Sel. Areas
  Commun.}, vol.~39, no.~11, pp. 3411--3424, Dec. 2021.

\bibitem{wuwei_UAVswarm_MEC}
W.~Wu, F.~Zhou, B.~Wang, Q.~Wu, C.~Dong, and R.~Q. Hu, ``Unmanned aerial
  vehicle swarm-enabled edge computing: Potentials, promising technologies, and
  challenges,'' \emph{IEEE Wireless Commun.}, vol.~29, no.~4, pp. 78--85, Aug.
  2022.

\bibitem{ACMM_FEI_2022}
B.~Fei, W.~Bao, X.~Zhu, D.~Liu, T.~Men, and Z.~Xiao, ``Autonomous cooperative
  search model for multi-{UAV} with limited communication network,'' \emph{IEEE
  Internet Things J.}, vol.~9, no.~19, pp. 19\,346--19\,361, Oct. 2022.

\bibitem{Multi_Zhao_2022}
N.~Zhao, Z.~Ye, Y.~Pei, Y.-C. Liang, and D.~Niyato, ``Multi-agent deep
  reinforcement learning for task offloading in {UAV}-assisted mobile edge
  computing,'' \emph{IEEE Trans. Wirel. Commun.}, vol.~21, no.~9, pp.
  6949--6960, Sep. 2022.

\bibitem{DUSF_WU_2022}
J.~Wu, C.~Luo, Y.~Luo, and K.~Li, ``Distributed {UAV} swarm formation and
  collision avoidance strategies over fixed and switching topologies,''
  \emph{IEEE Trans Cybern}, vol.~52, no.~10, pp. 1525--1539, Oct. 2022.

\bibitem{TAMW_Bai_2023}
Y.~Bai, H.~Zhao, X.~Zhang, Z.~Chang, R.~Jäntti, and K.~Yang, ``Towards
  autonomous multi-{UAV} wireless network: A survey of reinforcement
  learning-based approaches,'' \emph{IEEE Commun. Surv. Tutor.}, Oct. 2023,
  early access.

\bibitem{Lei_TICUS_2021}
L.~Lei, G.~Shen, L.~Zhang, and Z.~Li, ``Toward intelligent cooperation of {UAV}
  swarms: When machine learning meets digital twin,'' \emph{IEEE Netw.},
  vol.~35, no.~1, pp. 386--392, Feb. 2021.

\bibitem{CPPM_Chen_2022}
J.~Chen, C.~Du, Y.~Zhang, P.~Han, and W.~Wei, ``A clustering-based coverage
  path planning method for autonomous heterogeneous {UAVs},'' \emph{IEEE Trans.
  Intell. Transp. Syst.}, vol.~23, no.~11, pp. 25\,546--25\,556, Dec. 2022.

\bibitem{Xiao_BBSCM_2021}
W.~Xiao, M.~Li, B.~Alzahrani, R.~Alotaibi, A.~Barnawi, and Q.~Ai, ``A
  blockchain-based secure crowd monitoring system using {UAV} swarm,''
  \emph{IEEE Netw.}, vol.~35, no.~1, pp. 108--115, Feb. 2021.

\bibitem{Bansal_STTOA_2022}
G.~Bansal, V.~Chamola, N.~Ansari, and B.~Sikdar, ``Scalable topologies for
  time-optimal authentication of {UAV} swarms,'' \emph{IEEE Netw.}, vol.~36,
  no.~6, pp. 126--132, Nov. 2022.

\bibitem{UAVswarm_formationControl}
B.~Xin, Y.~Kang, and W.~Liu, ``Tracking-oriented formation control for unmanned
  aerial vehicle swarm through output feedback under jointly-connected
  topologies,'' \emph{IEEE Trans. Veh. Technol.}, Sep. 2023, early access.

\bibitem{Cognitive_Hilal_2023}
W.~Hilal, S.~A. Gadsden, and J.~Yawney, ``Cognitive dynamic systems: A review
  of theory, applications, and recent advances,'' \emph{Proc IEEE Inst Electr
  Electron Eng}, vol. 111, no.~6, pp. 575--622, Jun. 2023.

\bibitem{Haykin_CDSBC_2016}
S.~Haykin, P.~Setoodeh, S.~Feng, and D.~Findlay, ``Cognitive dynamic system as
  the brain of complex networks,'' \emph{IEEE J. Sel. Areas Commun.}, vol.~34,
  no.~10, pp. 2791--2800, Sep. 2016.

\end{thebibliography}

\end{document}